\documentclass[draft,grl]{agutex}
\usepackage[dvips]{graphicx}
\authorrunninghead{NINA AND \v{C}ADE\v{Z}}
\titlerunninghead{DETECTION OF AGW IN LOWER IONOSPHERE BY VLF WAVES}
\authoraddr{Corresponding author: A. Nina,
Institute of Physics, University of Belgrade, Pregrevica 118, 11080 Belgrade, Serbia.
(sandrast@ipb.ac.rs)}
\begin{document}
\title{Detection of acoustic-gravity waves in lower ionosphere by VLF radio waves}
\authors{A. Nina\altaffilmark{1}
 and V.~M.~\v{C}ade\v{z}\altaffilmark{2}}

\altaffiltext{1}{Institute of Physics, University of Belgrade, Pregrevica 118, 11080 Belgrade, Serbia.}

\altaffiltext{2}{Astronomical Observatory, Volgina 7, 11060 Belgrade, Serbia.}
\begin{abstract}
We present a new method to study harmonic waves in the low ionosphere (60 - 90 km) by detecting their effects on reflection of
very low frequency (VLF) radio waves. Our procedure is based on amplitude analysis of reflected VLF radio waves
recorded in real time, which yields an insight into the dynamics of the ionosphere at heights where VLF radio waves are being reflected.
The method was applied to perturbations induced by the solar terminator motions at sunrises and sunsets. The obtained results show that typical perturbation frequencies found to exist in higher regions of the atmosphere are also present in the
lower ionosphere, which indicates a global nature of the considered oscillations. In our model atmosphere, they turn out to be the acoustic and gravity waves with comparatively short and long periods, respectively.
\end{abstract}

\begin{article}
\section{Introduction}

The terrestrial atmosphere is a medium of a highly complex nature concerning  numerous physical,
chemical and geometrical features that govern its versatile dynamics on various spatial and temporal
scales. Among prominent signatures are harmonic and quasi-harmonic  hydrodynamic motions
(including soliton formation and vortices  \citep{jov01}) generated by perturbers located
in the atmosphere (electric discharges in lightnings \citep{ina10} and atmospheric convective motions
\citep{svin09}), in the lithosphere (tectonic motions \citep{dau09}), and in
the outer
space (sunrise and sunset effects \citep{afr09}, and solar wind gust impacts on the upper boundaries of the atmosphere \citep{dek01a,dek01b}).
 In addition to natural perturbers,
there are also those caused artificially like nuclear explosions \citep{yan12}.
The induced perturbations may result into various large amplitude
nonlinear structures \citep{ste10}, and patterns of
either eigen-modes or driven linear waves in the atmosphere. The latter represents a
unique response of the atmospheric medium and can be used as a means to recover
some of its local physical
characteristics. Such waves are detected and studied by various methods
utilizing satellites, and by numerous techniques from the ground
using radars, rockets and radio waves.

In what follows, we focus our attention on perturbations
induced in the lower ionosphere, below 90 km, at sunrises and sunsets due to motions of
the solar terminator (ST) when transitions between the daytime and nighttime
result into abrupt changes of thermal heating and ionization conditions and eventually
lead to generation of a variety of wave modes. These are
generally acoustic and gravity waves (AGWs) in practically nonmagnetized plasma.

The properties of AGWs generated by ST in higher ionospheric layers
(above 90 km) have been studied by computing the total electron contents (TEC)
from data obtained by GPS \citep{afr08},
Doppler sounding \citep{sin12} and incoherent scatter radar methods \citep{gal98}.
In our paper, we extend studies of the ST excited AGWs to the lowest ionospheric region
below 90 km. We apply a new procedure of detection
AGWs that is based on the Fourier analysis of amplitudes of very low frequency (VLF) radio
waves registered in real time.

The VLF signals are being emitted continuously in time
by a system of VLF transmitters distributed world-wide and, after being deflected
from the ionosphere at heights below 90 km, they
are recorded by networks of VLF receivers.
The registered VLF wave variations in
real time reflect the non-stationary physical and chemical conditions
in perturbations in the medium along the VLF wave trajectories
\citep{nin11,nin12b}.
Namely, the VLF signal reflection hight depends on local electron density, and it is time dependent due to presence of perturbations  \citep{nin12a}. This further causes time variations of the VLF wave trajectory and, consequently, variations of the registered wave amplitude and phase. The analysis of spectral composition of VLF wave changes in real time therefore allows us to make conclusions on atmospheric oscillation patterns.

\section{Observational data and their treatment}

Our analysis involves data on registered VLF amplitude changes
arising from the ST passage for
five days: May 1, 2, 9, 12, and 13, 2010. The reason for choosing
these particular days was relatively quiet conditions without significant
traveling ionospheric disturbance events resulting from auroral activity,
atmospheric lightnings, and solar flares among others.
The observed VLF signal is emitted by the transmitter DHO in Germany
at 23.4 kHz and registered by the Atmospheric Weather
Electromagnetic System for Observation Modeling and Education (AWESOME)
receiver system in Institute of Physics, Belgrade, Serbia (a part of
Stanford/AWESOME Collaboration for Global VLF Research).

Fig.~\ref{f1} shows
the recorded VLF signal amplitudes $A(t)$ in real time for selected four
time intervals of 90 minutes that cover periods before and after the
sunrise and sunset, respectively, when quasi stationary conditions
of the basic state are achieved (domains a and b, and c and d,
respectively). Such conditions then allow application of
standard normal mode analysis to excited small amplitude perturbations
whose detection and study are the main issue of this paper.
Shaded time intervals in Fig.~\ref{f1} that separate the quasi
stationary domains, indicate the occurrence of complex processes
due to ST that perturb the atmospheric layers and VLF wave reflection and
propagation conditions, which results into a highly non-stationary
behavior of the recorded VLF signal amplitude at sunsets and sunrises.
Particular features of these amplitude variations were studied elsewhere
(see \cite{car65,cli99}) and we take them as
a fact meaning that a quasi-stationary basic state is being perturbed during a given
period of time (around the sunrises and sunsets) and it eventually evolves
into a new different quasi-stationary basic state. Each of these two
quasi-stationary basic states contains its own characteristic oscillation
spectrum. Namely, the nighttime-daytime and
daytime-nighttime transitions cause perturbations and leave traces in the
atmospheric wave-dynamics in the region where the recorded VLF signals
propagate and reflect. Consequently, the recorded wave amplitude $A(t)$ suffers specific
time variations which, when Fourier transformed, yield  the  oscillation
spectrum $A_F(\omega)$ of hydrodynamic wave-modes typical of the locally perturbed
atmospheric structure:

\begin{equation}
\label{e01}  A_F(\omega)=\frac{1}{\sqrt{2\pi}}\int_{-\infty}^{+\infty}{\rm{e}}^{-i\omega t}A(t)dt,
\end{equation}
where $\omega\equiv 2\pi/\tau$ and $\tau$ are the oscillation frequency and oscillation
period, respectively.
These spectra contain a summary information on numerous time varying influences. Some of them are irregular and their effects are present non-periodically.
Others are more regular like the solar radiation which is dominant during the daytime, and cosmic radiation that is more evident at nighttime, for example.  All these effects together make the spectrum complex and different for each quasi stationary domain. The extraction of waves excited by the ST
can be carried out by considering the following typical characteristics of the phenomenon:
they appear during the ST, after that they are being attenuated to a certain degree, these processes occur both at sunrise and sunset in spite of different daytime/nighttime conditions of the medium, and all this together is repeating from day to day.

To see better the effect of induced perturbations, we compute and plot in Fig.~\ref{f2} the ratios of amplitudes:
\begin{equation}
\alpha_{ba}(\tau)\equiv \frac{A_F(\tau;b)}{A_F(\tau;a)},\mbox{    }
\alpha_{dc}(\tau)\equiv \frac{A_F(\tau;d)}{A_F(\tau;c)}
\label{e01a}
\end{equation}
related to time intervals after (the domains $b$ and $d$) and before
(the domains $a$ and $c$) the local passage of the ST
during the sunrise and sunset, respectively. These two ratios  are related to
perturbations present at different VLF wave reflection altitudes under the nighttime and daytime conditions,
and their local maxima indicate the perturbation frequency spectrum excited by the ST passage and some other possible
mechanisms arising from different altitude sampling, time dependent variations of solar radiation, etc.
The effect due to the ST alone can now be filtered out in two steps.
First, we introduce additional amplitude ratios:
\begin{equation}
\alpha_{bc}(\tau)\equiv \frac{A_F(\tau;b)}{A_F(\tau;c)},\mbox{    }
\alpha_{da}(\tau)\equiv \frac{A_F(\tau;d)}{A_F(\tau;a)}
\label{e01b}
\end{equation}
that compare  perturbation spectra
related to two different parts of both the daytime and nighttime section of the day (domains b and c, and
d and a in Fig.~\ref{f1}, respectively) which assumes practically equal altitude sampling for each of the
two time intervals. Coefficients given by Eq.~(\ref{e01b}) are ratios of Fourier amplitudes appearing
in the time interval at the beginning of the daytime/nighttime section to those from the end of the section,
which indicates spectral attenuation of excited waves. In other words, peaks in plots for
$\alpha_{bc}(\tau)$ and $\alpha_{da}(\tau)$ in Fig.~\ref{f2} mean that
corresponding perturbations have amplitudes that are much smaller at the end than at
the beginning of the related time section and, consequently, they can be taken as those excited by the ST at the same sampling altitude.
The second step of filtering ST effects out is taking data for more days into consideration which
enables notification of excitations by some other sporadic perturbers.

\section{The model and linear perturbations}
\label{res}

To understand the physical nature of excited hydrodynamic oscillation spectra one has to perform the normal mode analysis
assuming linear perturbations in a model atmosphere.

At heights below $90$ km where VLF
radio waves are being deflected a typical atmosphere model
gives $n_n\sim 10^{21}\rm{m}^{-3}$ for the neutral particle density and only $n_p\sim 10^{8}\rm{m}^{-3}$
for charged plasma particles meaning that here the electric and magnetic
effects play little role in local dynamics and that standard hydrodynamic equations can be applied.
Thus, we consider adiabatic waves in an ideal neutral gas assumed isothermal
and in hydrostatic equilibrium, whose horizontal wavelengths $\lambda_{x,y}$
are sufficiently small in comparison with the radius of the Earth $R_E=6371$ km while the vertical
wavelength $\lambda_z$ is sufficiently small in comparison with both $R_E$ and some
temperature inhomogeneity length.
This allows for application of the plane parallel geometry with gravitational acceleration $\vec g=-g\vec e_z$
($g$ = 9.81 m/s$^2$) in a locally isothermal medium, which eventually yields the dispersion relation \citep{yeh72}:

\begin{equation}
\label{e09}\omega^4-\left(k_0^2+k_z^2+\frac{1}{4H^2}\right)v_s^2\omega^2+k_0^2v_s^2N^2_{BV}=0.
\end{equation}
where: $k_{x,y,z}\equiv 2\pi/\lambda_{x,y,z}$; $k_0^2=k_x^2+k_y^2$ is the horizontal wavenumber;
$H=v_s^2/(\gamma g)$ is the characteristic scaleheight of isothermal atmosphere; $\gamma$ is the
standard ratio of specific heats; $v_s^2$ is the adiabatic sound speed squared;
and  $N{_{BV} }$ is the
Brunt-V\"ais\"al\"a frequency given as $N_{BV}^2=(\gamma-1)g^2/v_s^2$.
The dispersion relation Eq.~(\ref{e09}) is quadratic in $\omega^2$ which indicates the existence of two wave-modes
in the considered medium: the acoustic and gravity mode.

In further analysis, it is convenient to express
the dispersion relation Eq.~(\ref{e09}) in terms of wavelengths and wave periods
in the following way:
\begin{equation}
\label{e10}
\lambda_0^2(\tau)={\cal D}_0(\tau)\left[1+\frac{{\cal D}_2(\tau)}{\lambda_z^2-{\cal D}_2(\tau)}\right]
\end{equation}
with:
\[
{\cal D}_0(\tau)=
\frac{v_s^2\tau_{_0}^2(\tau_{_{BV}}^2-\tau^2)}{\tau_{_{BV}}^2(\tau_{_0}^2-\tau^2)}\tau^2, \quad
{\cal D}_2(\tau)=\frac{v_s^2\tau_{_0}^2}{\tau_{_0}^2-\tau^2}\tau^2,
\]
and $\tau_{_0}=4\pi v_s/g\gamma$, $\tau_{_{BV}}=2\pi/N_{BV}$ and $\lambda_{0,z}=2\pi/k_{0,z}$,
where $\tau_0$ and $\tau_{_{BV}}$ correspond to the acoustic cut-off and Brunt-V\"ais\"al\"a  oscillation period, respectively.

The dispersion relation Eq.~(\ref{e10}) represents a family of hyperbolas in a ($\lambda_0$,$\lambda_z$)-plot with $\tau$
being a parameter. This is shown in Fig.~\ref{f3} taking 250 K as typical
temperature of the ionosphere below 90 km.
Depending on $\tau$, Eq.~(\ref{e10}) has two separate domains describing
the acoustic modes if $0<\tau\le\tau_{_0}=$ 264.6 s, and gravity modes for $\tau\ge\tau_{_{BV}}=$ 270.0 s. There is no
propagating waves with $\tau_{_{BV}}\ge\tau\ge\tau_{_0}$.

\section{Discussion and conclusions}
\label{concl}

\begin{figure}
\noindent\includegraphics[width=7.5cm]{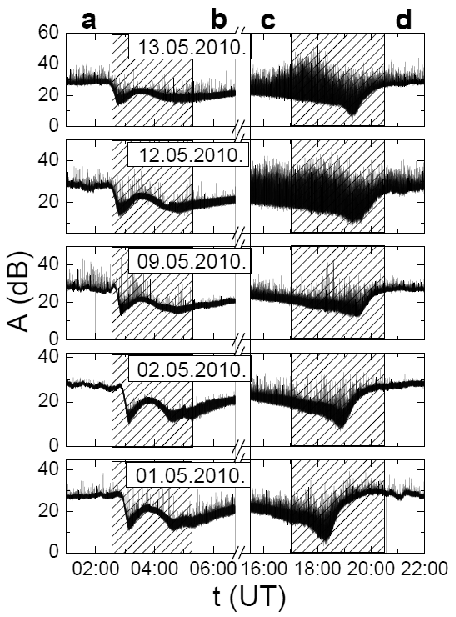}
\caption{Recorded VLF wave amplitude variations in real time for five indicated days
during periods of quasi stationary (preceding (a) and following (b) sunrises, and
preceding (c) and following (d) sunsets) and non stationary (shaded domains) basic state.}
\label{f1}
\end{figure}

\begin{figure}
\noindent\includegraphics[width=15.5cm]{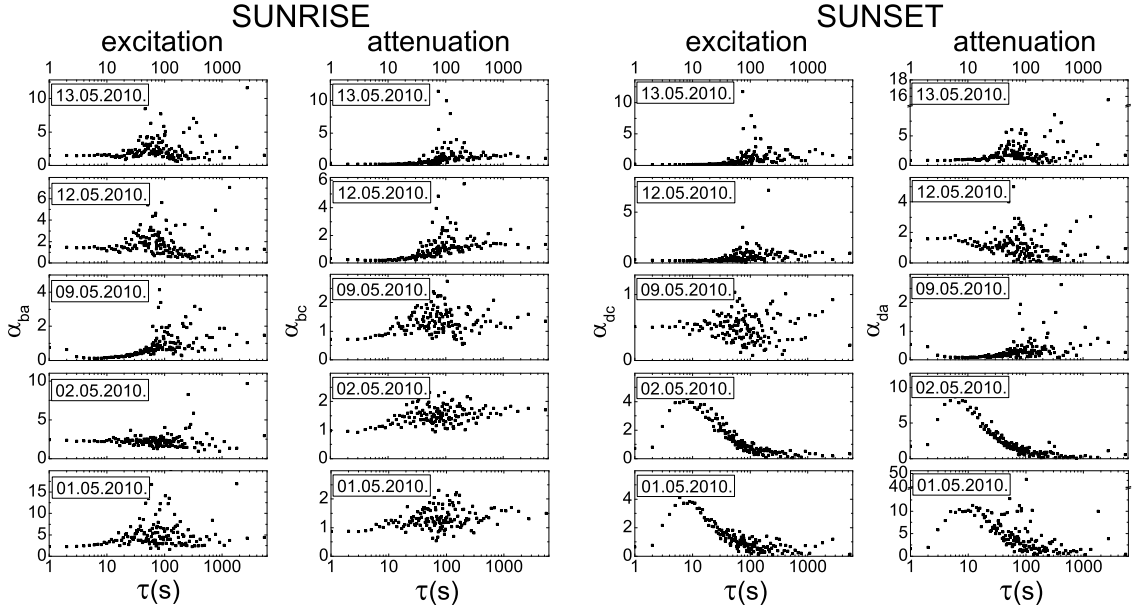}
\caption{Spectral amplitude ratios  $\alpha_{ba}(\tau)$ and
$\alpha_{bc}(\tau)$ related to sunrise, and spectral amplitude ratios  $\alpha_{dc}(\tau)$ and
$\alpha_{da}(\tau)$ related to sunset.
}
\label{f2}
\end{figure}

To visualize the effect of the ST passage on excitation of AGW modes, we plot the ratios $\alpha_{ba}(\tau)$ and
$\alpha_{dc}(\tau)$ (panels labeled "excitation" in Fig.~\ref{f2}) of
amplitudes of Fourier spectra for VLF waves recorded within the two quasi stationary time periods after and before the period of
ST induced disturbances (shaded domains in Fig.~\ref{f1}) both at the sunrise and sunset for five indicated days.
As these two spectra correspond to different reflection heights due to different local solar radiation conditions during the two time intervals, they may contain some spectral features of possible local origins in addition to AGW modes excited by ST.
 To identify the contribution of the ST passage, we compare spectral properties of perturbations excited at the same height by plotting
the ratios $\alpha_{bc}(\tau)$ and $\alpha_{da}(\tau)$ (panels labeled "attenuation" in Fig.~\ref{f2}) related to the quasi stationary time periods at the beginning and end of the daytime and nighttime, respectively. Consequently, these plots indicate changes in amplitudes, i.e. an expected wave damping in the span of daytime/nighttime for perturbations induced by the ST passage.

\begin{figure}
\noindent\includegraphics[width=7.5cm]{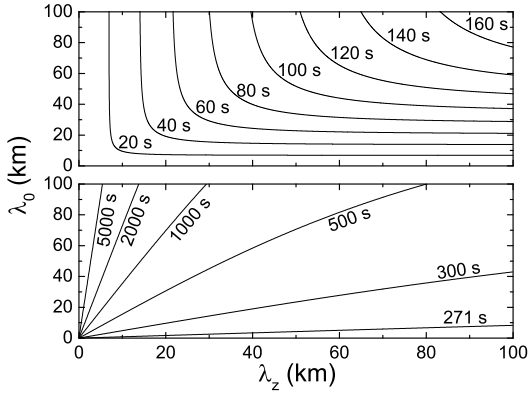}
\caption{The acoustic and gravity mode branches (the upper and lower panel, respectively)
of the dispersion relation Eq.~(\ref{e10}) for the considered isothermal
atmosphere with $T_0=250$ K.}
\label{f3}
\end{figure}

Fig.~\ref{f2} shows a daily repeating of relative amplitude enhancements for all considered ratios at several distinct oscillation periods within $\tau=$ 1 - 5400 s.  The most pronounced harmonic perturbations occur at periods $\tau\approx$ 60 - 100 s.
Going to Fig.~\ref{f3}, these oscillation periods correspond to
acoustic modes with the horizontal wavelength $\lambda_0\approx$ 20 - 40 km and for any
$\lambda_z>$ 20 km, or with any horizontal wavelength $\lambda_0>$ 20 km and $\lambda_z\approx$ 20 - 40 km.
Another noticeable ST excitations both at sunrises and sunsets occur at $\tau\approx$ 300 - 400 s and
at longer periods with $\tau\approx$ 1000 - 3000 s as seen in Fig.~\ref{f2}. These
perturbations are related to gravity modes with allowed wavelengths $\lambda_0$ and $\lambda_z$
as shown in Fig.~\ref{f3}.

In addition to the described properties of amplitude enhancement, there are also some isolated features
present at certain days only, like broad peaks around $\tau\approx$ 7 - 10 s
appearing on May 1 and 2, 2010 (Fig.~\ref{f2})
at sunsets, and for this reason they cannot be related to ST.
We also notice amplitude enhancements corresponding to
gravity modes with periods $\tau\approx$ 700 - 900 s present during sunrises while,
in most cases, they remain practically absent at sunsets. A similar
conclusion that ST-generated AGWs can be more pronounced at sunrise then in sunset was also obtained for heights above 100 km by \cite{afr08} using the TEC determined by GPS technics.

Our results for the low ionosphere are very similar to those for
higher altitudes of the E and F regions from literature.
Thus, excitation of AGWs  with periods $\tau<$ 1200 s is found to be
the main origin of the medium-scale traveling ionospheric disturbances
in the E and F regions \citep{her06}. Also, TEC oscillations with periods of about 15 min and those with 1 h were
reported in \cite{afr08}.
Similar fluctuations in form of MHD waves were also found
in high magnetospheric regions as externally driven modes with typical periods ranging
from few seconds to more than 1000 s \citep{dek01a,dek01b}.

To conclude, we present an extension of studies of the ST induced AGWs to altitudes of the low ionosphere (60 - 90 km).
Our procedure is based on the analysis of amplitudes of reflected VLF radio waves that are recorded in real time and it yields an insight into
the dynamics in the ionosphere at heights where the VLF waves are being reflected.
The obtained results show that the specific perturbation frequencies found to exist in higher regions of the atmosphere are also present in the low ionosphere. In our model atmosphere, they turn out to be the acoustic and gravity waves with comparatively short and long periods, respectively. In other words, the considered periodic perturbation motions in the atmosphere may be of a global
nature in a sense that they are attributed to different wave-modes of the same oscillation periods existing under physical conditions at corresponding locations in the atmosphere.

%

\begin{acknowledgments}
The present work was supported by the Ministry
of Education, Science and Technological Development of the Republic of Serbia as a part of the projects no. III 44002, 176002 and 176004.
\end{acknowledgments}

\bibliographystyle{agufull08}

\begin{thebibliography}{31}
\providecommand{\natexlab}[1]{#1}
\expandafter\ifx\csname urlstyle\endcsname\relax
  \providecommand{\doi}[1]{doi:\discretionary{}{}{}#1}\else
  \providecommand{\doi}{doi:\discretionary{}{}{}\begingroup
  \urlstyle{rm}\Url}\fi

\bibitem[{\textit{{Afraimovich}}(2008)}]{afr08}
{Afraimovich}, E.~L. (2008), {First GPS-TEC evidence for the wave structure
  excited by the solar terminator}, \textit{Earth Planets Space},
  \textit{60}, 895--900.

\bibitem[{\textit{{Afraimovich} et~al.}(2009)\textit{{Afraimovich}, {Edemskiy},
  {Leonovich}, {Leonovich}, {Voeykov}, and {Yasyukevich}}}]{afr09}
{Afraimovich}, E.~L., I.~K. {Edemskiy}, A.~S. {Leonovich}, L.~A. {Leonovich},
  S.~V. {Voeykov}, and Y.~V. {Yasyukevich} (2009), {MHD nature of night-time
  MSTIDs excited by the solar terminator}, \textit{Geophys. Res. Lett.}, \textit{36}, 15,106, \doi{10.1029/2009GL039803}.

\bibitem[{\textit{{Carpenter} and {Whitson}}(1965)}]{car65}
{Carpenter}, G.~B., and A.~L. {Whitson} (1965), {Observation of NPG VLF
  Transmissions at Tracy, California During Path Equinox}, \textit{J. Res.
  Natl. Bur. Stand., Sec. D: Radio Sci.}, \textit{69D}, 4621--4628,
  \doi{10.6028/jres.069D.074}.

\bibitem[{\textit{{Clilverd} et~al.}(1999)\textit{{Clilverd}, {Thomson}, and
  {Rodger}}}]{cli99}
{Clilverd}, M.~A., N.~R. {Thomson}, and C.~J. {Rodger} (1999), {Sunrise effects
  on VLF signals propagating over a long north-south path}, \textit{Radio
  Sci.}, \textit{34}, 939--948, \doi{10.1029/1999RS900052}.

\bibitem[{\textit{{Dautermann} et~al.}(2009)\textit{{Dautermann}, {Calais},
  {Lognonn{\'e}}, and {Mattioli}}}]{dau09}
{Dautermann}, T., E.~{Calais}, P.~{Lognonn{\'e}}, and G.~S. {Mattioli} (2009),
  {Lithosphere-atmosphere-ionosphere coupling after the 2003 explosive eruption
  of the Soufriere Hills Volcano, Montserrat}, \textit{Geophys. J. Int.}, \textit{179}, 1537--1546,
  \doi{10.1111/j.1365-246X.2009.04390.x}.

\bibitem[{\textit{{De Keyser} and {{\v C}ade{\v
  z}}}(2001{\natexlab{a}})}]{dek01a}
{De Keyser}, J., and V.~{{\v C}ade{\v z}} (2001{\natexlab{a}}), {Excitation of
  low-frequency fluctuations at the magnetopause by intermittent broadband
  magnetosheath waves}, \textit{J. Geophys. Res.}, \textit{106},
  29,467--29,478, \doi{10.1029/2001JA900078}.

\bibitem[{\textit{{De Keyser} and {{\v C}ade{\v
  z}}}(2001{\natexlab{b}})}]{dek01b}
{De Keyser}, J., and V.~{{\v C}ade{\v z}} (2001{\natexlab{b}}), {Transient
  development of magnetohydrodynamic wave mode conversion layers},
  \textit{J. Geophys. Res.}, \textit{106}, 15,609--15,620,
  \doi{10.1029/2001JA900045}.

\bibitem[{\textit{Galushko et~al.}(1998)\textit{Galushko, Paznukhov, Yampolski,
  and Foster}}]{gal98}
Galushko, V.~G., V.~V. Paznukhov, Y.~M. Yampolski, and J.~C. Foster (1998),
  Incoherent scatter radar observations of agw/tid events generated by the
  moving solar terminator, \textit{Ann. Geophys.}, \textit{16}(7),
  821--827, \doi{10.1007/s00585-998-0821-3}.

\bibitem[{\textit{{Hern{\'a}ndez-Pajares}
  et~al.}(2006)\textit{{Hern{\'a}ndez-Pajares}, {Juan}, and {Sanz}}}]{her06}
{Hern{\'a}ndez-Pajares}, M., J.~M. {Juan}, and J.~{Sanz} (2006), {Medium-scale
  traveling ionospheric disturbances affecting GPS measurements: Spatial and
  temporal analysis}, \textit{J. Geophys. Res.},
  \textit{111}, A07S11, \doi{10.1029/2005JA011474}.

\bibitem[{\textit{{Inan} et~al.}(2010)\textit{{Inan}, {Cummer}, and
  {Marshall}}}]{ina10}
{Inan}, U.~S., S.~A. {Cummer}, and R.~A. {Marshall} (2010), {A survey of ELF
  and VLF research on lightning-ionosphere interactions and causative
  discharges}, \textit{J. Geophys. Res.},
  \textit{115}, A00E36, \doi{10.1029/2009JA014775}.

\bibitem[{\textit{{Jovanovi{\'c}} et~al.}(2001)\textit{{Jovanovi{\'c}},
  {Stenflo}, and {Shukla}}}]{jov01}
{Jovanovi{\'c}}, D., L.~{Stenflo}, and P.~K. {Shukla} (2001), {Acoustic gravity
  tripolar vortices}, \textit{Phys. Lett. A}, \textit{279}, 70--74,
  \doi{10.1016/S0375-9601(00)00796-9}.

\bibitem[{\textit{{Nina} et~al.}(2011)\textit{{Nina}, {{\v C}ade{\v z}},
  {Sre{\'c}kovi{\'c}}, and {{\v S}uli{\'c}}}}]{nin11}
{Nina}, A., V.~{{\v C}ade{\v z}}, V.~A. {Sre{\'c}kovi{\'c}}, and D.~{{\v
  S}uli{\'c}} (2011), {The Influence of Solar Spectral Lines on Electron
  Concentration in Terrestrial Ionosphere}, \textit{Balt. Astron.},
  \textit{20}, 609--612.

\bibitem[{\textit{{Nina} et~al.}(2012{\natexlab{a}})\textit{{Nina}, {{\v
  C}ade{\v z}}, {{\v S}uli{\'c}}, {Sre{\'c}kovi{\'c}}, and {{\v
  Z}igman}}}]{nin12b}
{Nina}, A., V.~{{\v C}ade{\v z}}, D.~{{\v S}uli{\'c}}, V.~{Sre{\'c}kovi{\'c}},
  and V.~{{\v Z}igman} (2012{\natexlab{a}}), {Effective electron recombination
  coefficient in ionospheric D-region during the relaxation regime after solar
  flare from February 18, 2011}, \textit{Nucl. Instrum. Meth.B}, \textit{279}, 106--109,
  \doi{10.1016/j.nimb.2011.10.026}.

\bibitem[{\textit{{Nina} et~al.}(2012{\natexlab{b}})\textit{{Nina}, {{\v
  C}ade{\v z}}, {Sre{\'c}kovi{\'c}}, and {{\v S}uli{\'c}}}}]{nin12a}
{Nina}, A., V.~{{\v C}ade{\v z}}, V.~{Sre{\'c}kovi{\'c}}, and D.~{{\v
  S}uli{\'c}} (2012{\natexlab{b}}), {Altitude distribution of electron
  concentration in ionospheric D-region in presence of time-varying solar
  radiation flux}, \textit{Nucl. Instrum. Meth.B}, \textit{279}, 110--113, \doi{10.1016/j.nimb.2011.10.019}.

\bibitem[{\textit{Sindelarova et~al.}(2009)\textit{Sindelarova, Buresova, and
  Chum}}]{svin09}
Sindelarova, T., D.~Buresova, and J.~Chum (2009), Observations of
  acoustic-gravity waves in the ionosphere generated by severe tropospheric
  weather, \textit{Stud. Geophys. Geod.}, \textit{53}, 403--418,
  \doi{10.1007/s11200-009-0028-4}.

\bibitem[{\textit{{Sindelarova} et~al.}(2012)\textit{{Sindelarova}, {Mosna},
  {Buresova}, {Chum}, {McKinnell}, and {Athieno}}}]{sin12}
{Sindelarova}, T., Z.~{Mosna}, D.~{Buresova}, J.~{Chum}, L.-A. {McKinnell}, and
  R.~{Athieno} (2012), {Observations of wave activity in the ionosphere over
  South Africa in geomagnetically quiet and disturbed periods},
  \textit{Adv. Space. Res.}, \textit{50}, 182--195,
  \doi{10.1016/j.asr.2012.04.016}.

\bibitem[{\textit{{Stenflo} and {Marklund}}(2010)}]{ste10}
{Stenflo}, L., and M.~{Marklund} (2010), {Rogue waves in the atmosphere},
  \textit{J. Plasma Phys.}, \textit{76}, 293--295,
  \doi{10.1017/S0022377809990481}.

\bibitem[{\textit{{Yang} et~al.}(2012)\textit{{Yang}, {Garrison}, and
  {Lee}}}]{yan12}
{Yang}, Y.-M., J.~L. {Garrison}, and S.-C. {Lee} (2012), {Ionospheric
  disturbances observed coincident with the 2006 and 2009 North Korean
  underground nuclear tests}, \textit{Geophys. Res. Lett.},
  \textit{39}, L02103, \doi{10.1029/2011GL050428}.

\bibitem[{\textit{{Yeh} and {Liu}}(1972)}]{yeh72}
{Yeh}, K.~C., and C.~H. {Liu} (1972), \textit{{Theory of ionospheric waves}}, Academic Press, New York.

\end{thebibliography}


\end{article}

\end{document}